# Enhancement of magnetic ordering temperature and magnetodielectric coupling by hole doping in a multiferroic DyFe$_{0.5}$Cr$_{0.5}$O$_3$

Mohit K. Sharma[1], Tathamay Basu[2, #], K. Mukherjee[1,*] and E. V. Sampathkumaran[2]

[1]School of Basic Sciences, Indian Institute of Technology Mandi, Mandi 175005, Himachal Pradesh, India

[2]Tata Institute of Fundamental Research, Homi Bhabha Road, Colaba, Mumbai 400005, India

*Email: kaustav@iitmandi.ac.in

## ABSTRACT

We report the results of our investigations of magnetic, thermodynamic and dielectric properties of Ca substituted half-doped orthochromite, Dy$_{0.6}$Ca$_{0.4}$Fe$_{0.5}$Cr$_{0.5}$O$_3$. Magnetic susceptibility and heat capacity data bring out that this compound undergoes two antiferromagnetic transitions, one at ~ 132 and other at ~ 22 K. These values are higher than those of DyFe$_{0.5}$Cr$_{0.5}$O$_3$. This finding highlights that non-magnetic hole doping in form of Ca$^{+2}$, in the place of magnetic Dy$^{+3}$, tends to enhance magnetic transition temperatures in this half-doped orthochromite. We attribute it to possible change in the valence state of Cr/Fe-ion ions due to hole doping. Dielectric anomalies are also seen near the magnetic ordering temperatures indicating magnetodielectric coupling, which is confirmed by magnetic field dependent dielectric studies. The most notable observation is that magnetodielectric coupling strength gets significantly enhanced as compared to DyFe$_{0.5}$Cr$_{0.5}$O$_3$. The results reveal that it is possible to tune magnetoelectric coupling by hole doping in this system.

**INTRODUCTION**

Among the class of emerging materials in fundamental science research, multiferroics turned out to be exciting ones due to application potential in the area of spintronics, data storage and sensors [1-6]. In the past couple of decades, the compounds with perovskite structure $ABO_3$ (A = rare-earth, B = transition metals), well studied in the areas of colossal magnetoresistance and high-$T_c$ superconductivity, gained attraction for magnetoelectrics and multiferroics research as well. In this context, an emerging class of materials, half-doped orthochromites $AFe_{0.5}Cr_{0.5}O_3$, which belong to perovskite structure, have been extensively investigated in recent years [7, 8]. Presence of two magnetic-moment containing transitions metal ions in the perovskite makes such compounds interesting from magnetism angle and its coupled properties. In this article, we focus on $DyFe_{0.5}Cr_{0.5}O_3$ [9, 10], which has been shown to be a multiferroic, exhibiting significant magnetocaloric effect (MCE) with magnetoelectric (ME) coupling playing a role to enhance MCE [9]. We have recently found that a partial replacement of Dy by some other rare-earth ions leads to enhancement of both MCE and electric polarization [10]. Such chemical substitutions do not involve any electron or hole doping. On the other hand, there is no study demonstrating how the magnetic and dielectric properties of $DyFe_{0.5}Cr_{0.5}O_3$ evolve due to dilution of rare-earth site by nonmagnetic divalent cation, that is by hole-doping. This is the aim of the present investigation.

In this article, we present a detailed investigation addressing the magnetic, thermodynamics and dielectric properties of $Dy_{0.6}Ca_{0.4}Fe_{0.5}Cr_{0.5}O_3$. We specifically chose Ca for the aim of this investigation, as the ionic radii of $Ca^{2+}$ (1 Å) is nearly equal to that of $Dy^{3+}$ (0.91 Å). Therefore, for the same crystal structure, a minimal change in chemical pressure can be obtained with such a 'hole' doping. It is therefore of interest to explore the result of such a hole doping on the dielectric or/and ferroelectric as well as on magnetic and magnetoelectric coupling properties. We would like to mention that the maximum solubility of Ca in $DyFe_{0.5}Cr_{0.5}O_3$ is about 40% of Dy, as revealed by our investigations. $DyFe_{0.5}Cr_{0.5}O_3$ has been found to undergo antiferromagnetic transitions at ~ 261, 121 and 13 K [9, 10]. In this manuscript, we focus on the magnetic transitions around 13 and 121 K only, as, the 261 K transition is so weak [10] that we cannot make any guess due to hole doping. It is observed that the lower and intermediate transition-temperatures are enhanced due to partial replacement of $Dy^{+3}$ by $Ca^{+2}$. Dielectric anomalies are seen at the magnetic ordering temperatures in $Dy_{0.6}Ca_{0.4}Fe_{0.5}Cr_{0.5}O_3$. Most interestingly, the magnetodielectric coupling strength also gets significantly enhanced with respect to that in the parent composition. Incidentally, our preliminary studies for a lower Ca content, namely for $Dy_{0.8}Ca_{0.2}Fe_{0.5}Cr_{0.5}O_3$, also indicated enhancement of magnetic transition temperature to ~ 126 K and ~ 19 K. Therefore, we decided to perform detailed studies on a composition with highest possible doping.

**EXPERIMENTAL DETAILS**

The polycrystalline sample, $Dy_{0.6}Ca_{0.4}Fe_{0.5}Cr_{0.5}O_3$ (DCFCO) was prepared by solid state reaction of stoichiometric amounts of $Dy_2O_3$, $CaCO_3$, $Fe_2O_3$ and $Cr_2O_3$ (Sigma-Aldrich, purity > 99.9%), under similar conditions as described in Ref. [8]. X-ray diffraction studies on the sample established that the specimen is crystallographically single phase within the detection limit of the technique and the pattern was also analyzed by the Rietveld profile refinement [11]. This analysis establishes that the sample crystallizes in the orthorhombic structure with the space group *Pbnm*. The lattice constants obtained from this analysis are a = 5.2869 (1) Å, b = 5.5089 (8) Å and c = 7.5734 (1) Å. It is observed that change in the lattice parameter is ≤1% as compared to $DyFe_{0.5}Cr_{0.5}O_3$ [10], implying that the lattice perturbation caused by Ca substitution can be neglected. Temperature ($T$) and magnetic field ($H$) dependent DC magnetization ($M$) data were collected by the Magnetic Property Measurements System (MPMS) from Quantum design, USA. Heat capacity (C) measurements were performed using Physical Property Measurements System (PPMS) from Quantum design, USA. Complex dielectric permittivity was measured as a function of temperature and magnetic field in the frequency range 30 to 500 kHz with 1V ac bias during warming (1 K/min) using Agilent E-4980A LCR meter with a homemade sample holder integrated with PPMS.

**RESULTS AND DISCUSSION**

The temperature response of dc magnetic susceptibility ($\chi=M/H$) obtained under the zero-field-cooling (ZFC), and field-cooled-cooling (FC) condition in the $T$ range from 2 to 150 K at 100 Oe applied magnetic field is shown in Fig. 1(a). It is observed that, under both the conditions, the magnetization increases as $T$ decreases up to ~ 24 K. On further decrease in $T$, $M$ is seen to decrease. The compound undergoes antiferromagnetic orderings, one at $T_1$ ~ 132 K and the other at $T_2$ ~ 22 K, as evident from the $d\chi/dT$ vs. $T$ plots which exhibit a peak at the respective transition temperatures shown in upper inset for $T_1$ and in the lower inset for $T_2$ in Fig 1(a). Thus, a partial replacement of Dy by Ca shifts $T_1$ and $T_2$ to the higher $T$ side, as compared to that of $DyFe_{0.5}Cr_{0.5}O_3$ [see a curve for the same, added from ref [10] for comparison], in which the magnetic ordering at $T_1$ is ascribed to Cr-O-Cr magnetic interactions and the one at $T_2$ to Dy-O-Fe/Cr magnetic interactions [9]. To understand the magnetic ordering behaviour in this compound, the temperature response of heat capacity is plotted as a function of $T$ from 2 to 150 K in the presence of different externally applied magnetic fields (Fig 1(b)). A weak anomaly peak is observed around $T_2$, but such a feature around $T_1$ is absent in the raw data. In order to see the features more clearly, the temperature response of $dC/dT$, is plotted in the inset. The derivative curve shows peaks around $T_1 \sim 131$ K (upper inset of Fig. 1(b)) and $T_2 \sim 20$ K (lower inset of Fig. 1(b)). Additionally, the peak temperature around $T_2$ is seen to decrease under an application of the magnetic field of 5 kOe (lower inset of Fig. 1(b)). Such a behaviour, that is, the decrease of the peak temperature for an application of $H$ viewed together with the feature observed in FC magnetization curve, where $M$ is seen to decrease below transition $T$, is characteristic of antiferromagnetic ordering. The observed enhancement is interesting, as the non-magnetic $Ca^{2+}$ should weaken the Dy-O-Fe/Cr the magnetic interaction. Therefore, some factors must be playing an important role to enhance exchange interaction. It is to be noted that when trivalent $Dy^{+3}$ ions are replaced by divalent $Ca^{+2}$ ions in $DyFe_{0.5}Cr_{0.5}O_3$, some $(Fe/Cr)^{+3}$ ions may change into $(Fe/Cr)^{+4}$ due to the charge imbalance at A and B sites in the perovskite $ABO_3$ structure [12], though we cannot rule out a change in the oxygen content. Clearly such imbalances play an important role to determine the physical properties in this system. Similar increase in transition temperatures due to Ca doping has also been observed in other transition metal oxides [13]. It may be stated that the temperature-dependent features due to magnetic transitions are sufficiently broad and non-hysteretic that the transitions are not first-order in character.

To get a better insight about the magnetic behavior of this compound, isothermal $M$ was measured as a function of $H$ (in the range ±50 kOe) at different $T$ (2 to 150 K). The $M(H)$ curves at different temperature (2, 40, and 150 K) are shown in Fig.1(c). A weak magnetic hysteresis is observed (see the inset of Fig. 1(c)) at low fields (≤ 5kOe) and the magnetization does not saturate at high fields. The non-saturation tendency of magnetization in the magnetically ordered state is generally noted for systems with finite antiferromagnetic coupling, whereas the presence of magnetic hysteresis is an indication of the presence of ferromagnetic component. The coercive field ($H_C$) exhibits a continuous increase up to 80 K with decreasing temperature and then it shows a decreasing trend (display in Fig. 1(d)). It is possible that there is a change in the relative alignment of transition-metal and rare-earth sub-lattice moments with $T$, which can also result in ferromagnetic and antiferromagnetic components [14, 15].

In order to see the effect Dy-site dilution by $Ca^{2+}$ on isothermal magnetic entropy change ($\Delta S_M$), $\Delta S_M$ was calculated from $M(H)$ isotherms [16]. Inset of Fig 1(d) shows the $T$ response of $|\Delta S_M|$ at 50 kOe. $\Delta S_M$ shows a broad peak and reaches a maximum value of ~ 6.3 J/kg-K around 10 K for $\Delta H$ = 50 kOe. It is observed that the peak temperature of $\Delta S_M$ does not match with $T_2$. The peak temperature of magnetic entropy change is same (~9 K) for both $DyFe_{0.5}Cr_{0.5}O_3$ [10] and this compound. Thus, it can be said that Ca-doping only affects the value of entropy change, but does not shift the peak temperature. This indicates that magnetic entropy change is not simply induced by magnetic ordering of rare-earth and transition metal ions and some other mechanism like magneto electric coupling (as also observed for these types of orthochromites [9]) is responsible for the observed behaviour. Similar behaviour of $\Delta S_M$ was also observed due to Ho doping on $DyMnO_3$, where the peak temperature of $\Delta S_M$ does not change with Ho doping [17].

To probe the dielectric behavior of the sample, we measured the temperature response of complex dielectric permittivity with different frequencies (10 to 500 kHz) in the temperature range 2 to 150 K in zero magnetic field. The $T$ responses of dielectric permittivity, both real ($\varepsilon'$) and loss part ($\tan\delta$) of the compound, are shown in Fig. 2(a) and (b) for $H=0$. No visible sharp change of features is observed in $\varepsilon'$ around $T_1$ or $T_2$; however, a change in slope is observed. The temperature derivative of $\varepsilon'$ shows anomalies around 22 K (lower inset of Fig. 2(a)) and 125 K (upper insets of Fig. 2(a)), these temperatures are nearly equal to the magnetic ordering temperatures. With respect to the parent composition, a strong frequency dependent shoulder is observed in the T-range from 70 to 120 K. Correspondingly, $\tan\delta$ shows a well-defined frequency dependent peak in the $T$ range 50-100 K, with peak temperature increasing with frequency. Such features are not seen in the parent composition. We attribute this to the development of Maxwell-Wagner (M-W) effect due to Ca-doping. That is, Ca substitution has a tendency to make the material less insulating and this argument is endorsed by the significantly enhanced values of $\tan\delta$ with respect to the composition without Ca. Alternatively, there could be a possibility of glassy dynamics (superimposed over dominating M-W behavior), as a result of the compositional fluctuations due to $Ca^{2+}$ ions substitution at the site of the perovskite structure [18]. We have not looked for the ferroelectric behaviour around this temperature region (which is present in $DyFe_{0.5}Cr_{0.5}O_3$) due to less insulating nature of the Ca containing composition. However, it is to be noted that the sample is highly insulating below 50 K, as inferred from very low value of tanδ (<<1). Therefore, we mainly restrict investigations of magneto-dielectric (MDE) effect to such a lower temperature range.

Fig 2(c) shows $T$ response of $\varepsilon'$ measured in the presence of different magnetic fields with 50 kHz. It was observed that $\varepsilon'$ increases with magnetic-field at temperatures below $T_1$, beyond which the effect of magnetic field is negligible. Hence, to investigate the nature of MDE, $\varepsilon'$ was measured as a function of $H$ at a few selected temperatures of 3, 10, 27 and 125 K with 50 kHz. The results of such measurements are shown in Fig 2(d), in the form of $\Delta\varepsilon'$ vs. $H$, where $\Delta\varepsilon' = [(\varepsilon'_H - \varepsilon'_{H=0})/\varepsilon'_{H=0}]$. The observed $\Delta\varepsilon'$ is positive and it increases with an increase in $H$. The value of $\Delta\varepsilon'$ at 3 K and at 50 kOe is ~ 3%. It reduces to half the value as the temperature is increased to 10 K and, above the magnetic ordering $T_2$, say at 27 K and above, its value reduces to nearly zero. Both the magnetization (see Fig. 1(c)) and $\Delta\varepsilon'$ for a fixed magnetic field decrease as $T$ increases consistent with MDE coupling [19]. In a nutshell, the observed magnitude of MDE at low temperatures is significantly increased (~ 15 times) by substituting non-magnetic divalent Ca-ion in the place of magnetic trivalent Dy-ion, while comparing with the behavior of the parent compound $DyFe_{0.5}Cr_{0.5}O_3$ (inset of Fig. 2d) [10]. It implies that, in the composition under investigation, the coupled spin - electric charge state is modified with Ca substitution in such a way that there is an increase in the coupling strength. The coupling strength is comparable (or even higher) to many well-known frustrated and spin-chain multiferroic compounds [19, 20]. We would also like to point out that the observed MDE coupling is intrinsic property of this compound, as the compound is highly insulating in the magnetically ordered region. The value of the electrical resistivity is ~800MΩ cm at 85 K. Below this temperature, resistance keeps increasing reaching GΩ range and hence could not be measured with proper resolution at further low temperatures. Also, no magnetoresistance was observed in the whole temperature range of measurement (inset of Fig. 2(c)).

**CONCLUSION**

In summary, a partial replacement of Dy by Ca in $DyFe_{0.5}Cr_{0.5}O_3$ results in the enhancement of the two magnetic transition temperatures, establishing the effect of hole doping on the magnetism in this system. Dielectric anomalies are seen near the magnetic ordering temperatures due to MDE coupling. MDE coupling as indicated by the magnetocapacitance values is significantly enhanced compared to that in the parent compound, thereby offering a channel to tune the coupling between electronic charge and spins by site dilution in these materials.

**Acknowledgments**

The authors acknowledge Kartik K. Iyer and V. Chandragiri for their help during experimental work. MKS and KM acknowledge experimental facilities of Advanced Material Research Centre (AMRC), IIT Mandi and also the financial support from IIT Mandi.

[#]Present Address: Experimental Physics V, Center for Electronic Correlations and Magnetism, University of Augsburg, D-86135 Augsburg, Germany

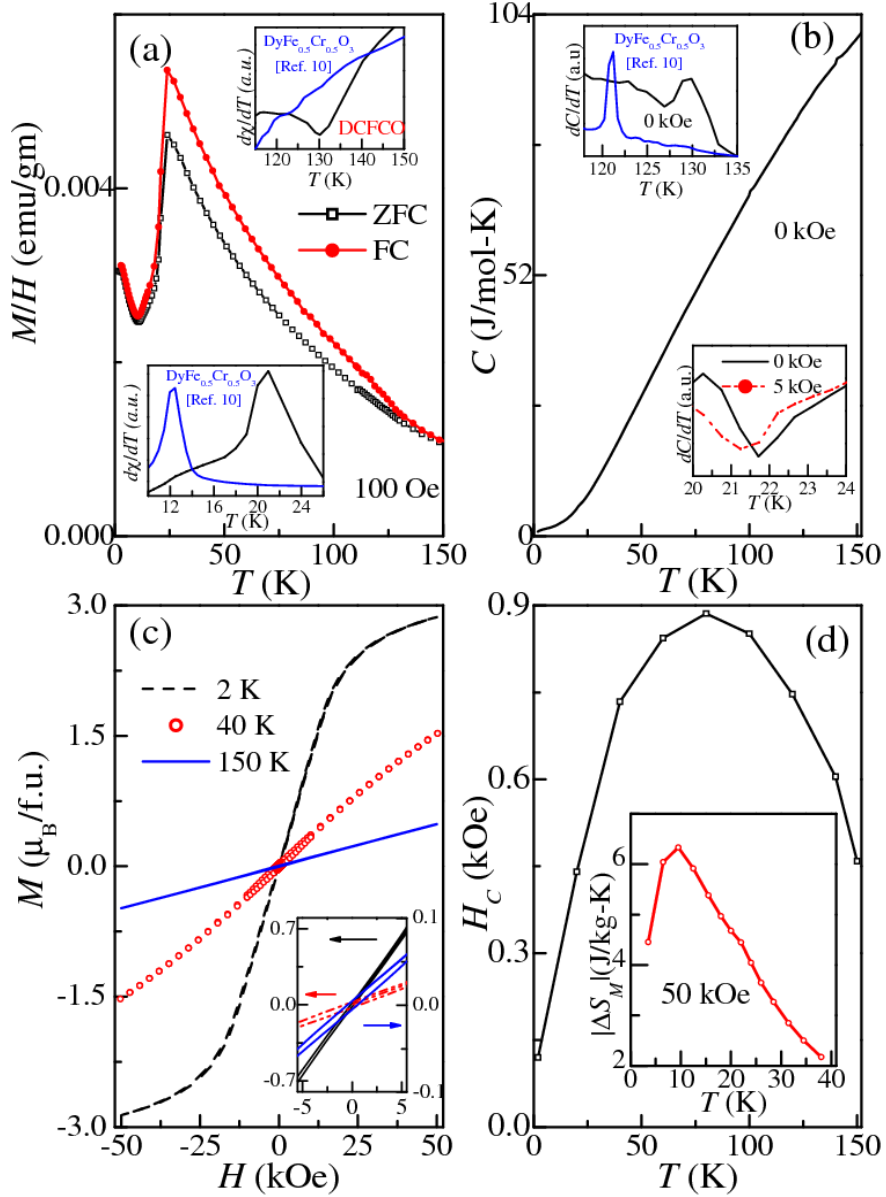

FIG. 1. (a) $T$ dependent ZFC and FC magnetization curve at 100 Oe. Upper inset: $d\chi/dT$ vs. $T$ plot in the range 115 to 150 K. Lower inset: $d\chi/dT$ vs. $T$ plot in in the range 10 to 25 K. Blue curve is for $DyFe_{0.5}Cr_{0.5}O_3$ from Ref 10. (b) $T$ response of $C$ at 0 kOe. Upper inset: $dC/dT$ vs. $T$ plot in the $T$ range 115 to 135 K at 0 kOe. Blue curve is for $DyFe_{0.5}Cr_{0.5}O_3$ from Ref 10. Lower inset: $dC/dT$ vs. $T$ plot in the $T$ range 20 to 24 K at 0 kOe and 5 kOe. (c) Isothermal $M$ ($H$) curve in the range ±50 kOe at different $T$ (2, 40 and 150 K). Inset: Magnified magnetic hysteresis in the range ±5 kOe. (d) $T$ dependent coercive field plot. Inset: Isothermal magnetic entropy change at 50 kOe.

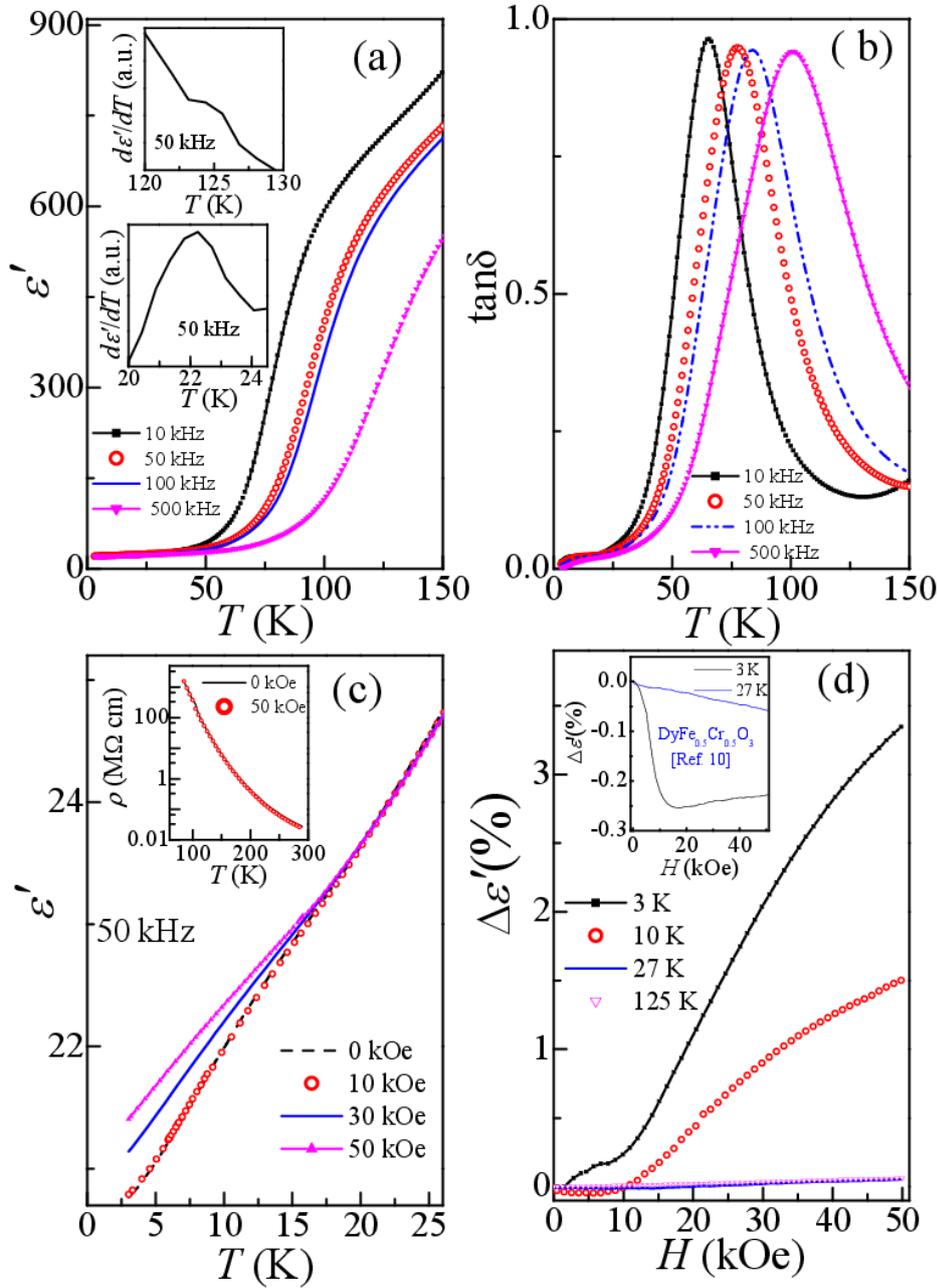

FIG. 2. (a) The plot of zero-field real part $\varepsilon'$ of dielectric permittivity vs. $T$ at selected frequencies (10, 50, 100 and 500 kHz). Upper inset: $d\varepsilon'/dT$ vs. $T$ plot in the temperature range of 120 to 130 K. Lower inset: $d\varepsilon'/dT$ vs. $T$ plot in the range of 20 to 24 K. (b) Zero field tan$\delta$ vs. $T$ plot at selected frequencies (10, 50, 100 and 500 kHz). (c) $\varepsilon'$ vs. $T$ plot at different magnetic fields (0, 10, 30 and 50 kOe). Inset: $T$ response of $\rho$ curve in zero and 50 kOe magnetic field. (d) Relative change in dielectric constant ($\Delta\varepsilon'$) by the application of external magnetic fields different temperature (3, 10, 27 and 125 K). Inset: $\Delta\varepsilon'$ vs. $H$ plot for $DyFe_{0.5}Cr_{0.5}O_3$ from Ref 10.